\documentclass[twocolumn,showpacs,amsmath,amssymb,aps,pra,nobalancelastpage,reprint]{revtex4}
\usepackage{amsthm}
\usepackage{newlfont}
\usepackage{graphicx}
\usepackage{epstopdf}
\usepackage{appendix}
\usepackage{hyperref}
\usepackage{textcomp}
\usepackage{appendix}
\usepackage{multirow}	
	
\begin{document}
\title{\textbf{Pinning of Hidden Vortices in Bose-Einstein Condensate}}
\author{T. Mithun$^1$, K. Porsezian$^{1}$ and Bishwajyoti Dey$^2$\\
\small \textsl{$^{1}$Department of Physics, Pondicherry University, Puducherry 605014, India.\\
$^2$Department of Physics, University of Pune, Pune 411007, India.}}
\begin{abstract}
We study the vortex dynamics and vortex pinning effect in Bose-Einstein condensate in a rotating double-well trap potential and  co-rotating optical lattice. We show  that,  in agreement with the experiment,  the vortex number do not diverge when the rotational frequency  $\Omega \rightarrow 1$ if the trap potential is  of anisotropic double-well type. The critical rotational frequency as obtained from numerical simulations agrees very well with the value $\sqrt l/l$ for $l=4$ which supports the  conjecture  that surface modes with angular momentum $l=4$ are excited when the rotating condensate is trapped in double-well potential. The vortex lattice structure in a rotating triple-well trap potential and its pinning shows very interesting features. We show the existence and pinning of a new type of hidden vortices whose phase profile is similar to that of the visible vortices.
\end{abstract}
\maketitle
\pagestyle{myheadings}
   One of the most striking properties of the Bose-Einstein condensate (BEC) is its ability to form  vortices and  vortex lattice in the rotating frame which manifest the superfluidity of BEC \cite{Matthews}. Inspired by the ability of ultra cold gases in optical lattice to explore a wide range of fundamental problems in condensed matter physics \cite{Greiner}, extensive theoretical and experimental studies have taken place with BEC in rotating optical lattice showing some interesting properties \cite{Polini}. Pinning of vortices and the associated the structural phase transition from the Abrikosov vortex lattice structure to the lattice structure of the optical lattice is the one of those properties \cite{Tung, Sato, Williams, Kato,Pu,Kasamatsu}.
\par
  The BEC in double-well potential trap have got much attention because of its richness in physics \cite{Albiez,Xiong, Juli, Spekkens, Salgueiro}. This motivated the study of the BEC trapped in rotating double-well potential \cite{Hofferberth}. Recent theoretical study about the Feynman rule in  BEC trapped in a rotating double well potential have unearthed the existence of hidden vortices \cite{Wen1}. These hidden vortices do not have visible cores but carries angular momentum. It has been shown that Feynman rule for the number of vortices in the condensate can be satisfied only after including the hidden vortices \cite{Wen1, Wen2}.
    \begin{figure}[pb]
       \vspace{-20pt}
  \includegraphics[width=8.6cm,height=4cm]{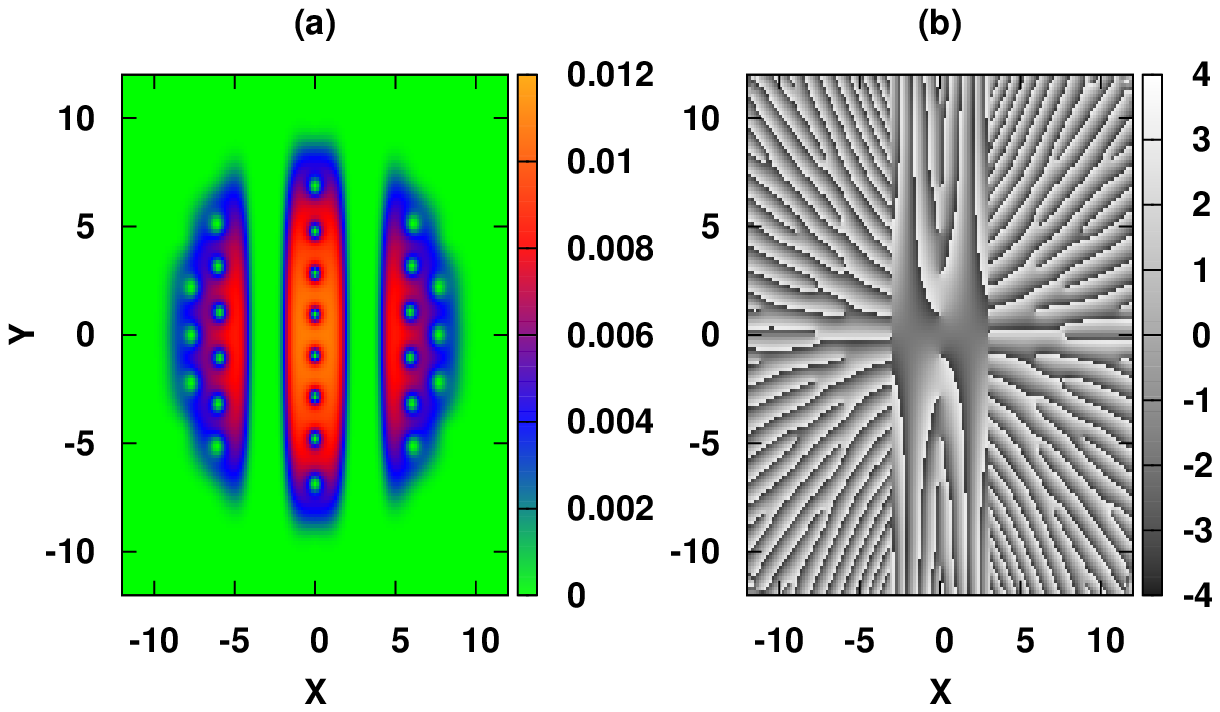}
     \vspace{-20pt}
 \caption{\label{Fig. 1}{\footnotesize (Color online) (a) Condensate density $|\psi^2|$ for triple-well trap and (b) phase profile of $\psi$ for $\Omega=0.9$ at t=300. Here $V_{0}=40$ and $\sigma=0.5$.}}
      \vspace{-20pt}
  \end{figure}
\par
   In this rapid communication, we study about the vortex formation and vortex pinning in BEC in a rotating double-well trap potential and co-rotating optical lattice. Our study is motivated by the observation of hidden vortices in BEC in a rotating double-well trap potential. It is natural to ask whether the hidden vortices can be pinned by the optical lattice.  Interestingly, our numerical simulations show that there exist two types of hidden vortices with different phase profiles and only the type which displays phase profile similar to that of the visible vortices can be pinned to the optical lattice.  Since the hidden vortices  are distributed along the central barrier region of the double-well trap potential, we have considered the case of triple-well trap potential where there are two barrier regions. In this case the nature of vortex lattice and its pinning effect shows very interesting features.  Another motivation for the present study is to derive the  well-known Feynman rule for the double-well trap potential\cite {Feynman}. It has been shown that for a  condensate trapped in a single-well (harmonic) potential and confined in a co-rotating optical lattice, the number of vortices increases linearly with rotational frequency $\Omega$ and diverges when the rotational frequency approaches the harmonic trap frequency \cite{Fetter}. Contrary to this, no such divergence in the vortex number have been observed in the recent experiments of Williams {\it et al} \cite{Williams}. Recently Kato {\it et al} \cite{Kato} argued that such divergence in the number of vortices can be avoided if one consider an optical lattice with Gaussian envelope of the laser beams \cite{Kato}. We, on the other hand show that if the trap potential is of double-well  anisotropic type, then there is no such  divergence in the number of vortices as rotational frequency approaches the trap frequency which is in agreement with the experiment.  We have also addressed the problem of the relation between the surface mode frequency and the critical rotational frequency $\Omega_c$ for rotating BEC in double-well trap potential. For the case of harmonic trap (single well) potential the initial surface mode excitation which leads to single vortex formation have angular momentum $l=2$.
 However, for the case of  double-well trap potential, the problem is nontrivial as in this case it is conjectured that higher order surface modes with \textit{l} = 4 are expected to contribute to initial turbulent motion of the rotating condensate which leads to the vortex formation \cite{Wen2}.
   \begin{figure}[pt]
          \vspace{-5pt}
 \includegraphics[width=8.6cm,height=4.7cm]{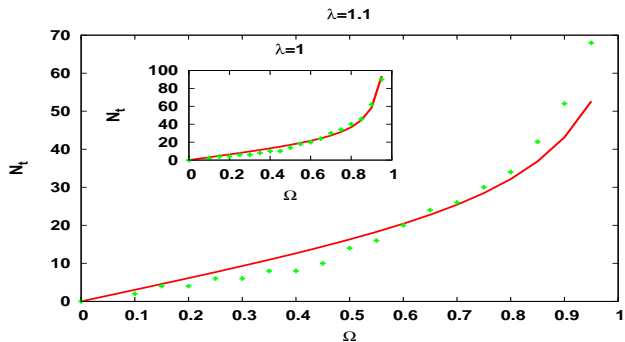}
   \vspace{-16pt}
 \caption{\label{Fig. 2}{\footnotesize (Color online) Total number of vortices Vs rotation frequency for isotropic case. The solid line represents analytical results, the points represent the numerical results.}}
  \vspace{-15pt}
 \end{figure}
    \begin{figure}[pht]
  \includegraphics[width=8.6cm, height=4.7cm]{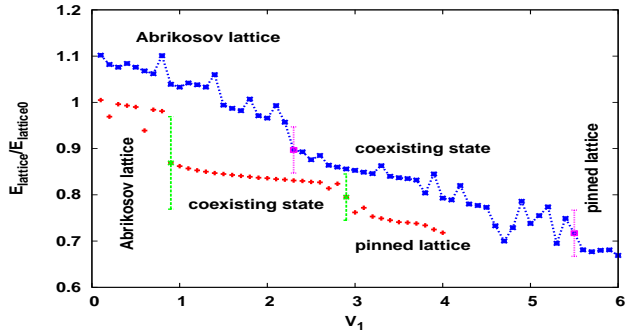}
   \vspace{-21pt}
 \caption{\label{Fig. 3}{\footnotesize (Color online) Lattice potential energy against the strength $V_{1}$ of the TOL (red dots) and SOL (blue line-dots). Here $\Omega=0.7$, $V_{0}=40$ and $\sigma=0.5$.}}
   \vspace{-15pt}
 \end{figure}
\par
We consider the two-dimensional (2D) dimensionless time-dependent Gross-Pitaevskii equation (GPE) $\small (i-\gamma)\psi_{t}=[-\frac{1}{2}(\bigtriangledown_{x}^2+\bigtriangledown_{y}^2 )+V(x,y)-\mu+p|\psi|^2-\Omega L_{z}]\psi$ as in reference \cite{Sato} for the formulation of our problem. Here the potential $V(x,y)$ is the sum of two potentials $V_{dw}(x,y)+V_{lattice}(\textbf{r})$. $V_{dw}(x,y)$  is the anisotropic double-well trap potential given by $\small V_{dw}(x,y)=\frac{1}{2}(x^2+ \lambda^2 y^2)+V_{0}e^{-x^2/2\sigma^2}$,
where $V_0$ denote the depth of the double well potential and $\lambda=\omega_{y}/\omega_{x}$ denote the anisotropy parameter. $V_{lattice}$ denote the optical lattice potential \cite{Sato,lattice}. The strength of the potentials $V_{0}$ and $V_{1}$ are in units of $\hbar \omega_{x}$.

The Crank-Nicolson scheme is used to numerically solve the 2DGPE.  For our simulation, we set small spatial step $\Delta x$ = $\Delta y$ = 0.04 and time step, $\Delta t$ = 0.0005. The parameters are chosen from the experiments of $^{87}Rb$ \cite{Williams,Kato}. The dissipation parameter is set to $\gamma =0.03$. The lattice constant $a$ is fixed as $a=2.2$ and $p =1000$.

 We obtain below the expression for the surface mode frequency for the rotating condensate trapped in a double-well potential using the time-dependent variational analysis. For this we set $V_{lattice}(\textbf{r})$=0 and $\lambda=1$. As the condensate is trapped in the double-well shaped potential, we use the ansatz for the condensate wave function to be of the form $\small \psi(x,y,t)=c(t) x^2 e^{\frac{-1}{2}[\alpha(t)x^2+\beta(t)y^2-2 i \gamma(t) x y]}$.
 In the presence of rotation, the centrifugal term $(-\Omega L_z)$ shifts the surface mode with $l=4$ (for the double-well trap potential) by $\pm 4\Omega$. By following the similar procedure for finding the surface modes frequency for harmonic trap potential case \cite{Mithun}, we obtain the lowest energy surface mode frequency for BEC trapped in a rotating double-well potential as  $\small \omega_{-4} = \omega - 4 \Omega$ \cite{Relation}.   This relation shows that the dynamical instability which leads to the visible vortex formation begins at $\Omega = \omega/4$.  Table I shows the variation of the surface mode frequency $\Omega = \omega/4 $ and the critical rotational frequency $\Omega_{c}$ obtained numerically with nonlinear interaction parameter $p$ and the depth of the double-well potential $V_0$.  It can be seen that $\Omega_{c}$ agrees quite well with the value ${\sqrt l}/l = 0.5$ for $l=4$. This shows  that the initial turbulent motion which leads to the vortex formation indeed  consist of the higher order surface modes with \textit{l} = 4. The comparison between the two frequencies shows that the surface mode frequency is much smaller than $\Omega_{c}$. This is unlike the harmonic (single well) trap case, where the surface mode frequency is quite close to $\Omega_{c}$ \cite{Mithun}.
       \begin{center}
   \vspace{-20pt}
    \begin{table}[t]\footnotesize
    \caption{Surface mode frequency $\Omega$ and critical rotational
frequency $\Omega_c$}
            \begin{tabular}{ p{1cm} |p{1.5cm} p{1.5cm} |p{1.5cm} c}
\hline
\multirow{2}{*}{p} & \multicolumn{2}{|l|}{Theoretical $\Omega=\omega/4$}& \multicolumn{2}{|l}{Numerical $\Omega_{c}$} \\ [0.5ex] \cline{2-5}
 & $V_{0}=40$ & $V_{0}=20$ & $V_{0}=40$ & $V_{0}=20$   \\ [0.5ex] \hline
   1000 &  0.258     & 0.220    &  0.47     & 0.46 \\
   800  &  0.279     & 0.237    &  0.49     & 0.48  \\
   600  &  0.307     & 0.260    &  0.52     & 0.52  \\
   400  &  0.346     & 0.297    &  0.57     & 0.56 \\
   200  &  0.402     & 0.359    &  0.62   & 0.63  \\ [1ex]
   \hline
\end{tabular}
     \end{table}
    \end{center}
We consider the vortex lattice formation in BEC trapped in symmetric triple-well potential given by  $\small V_{tw}(x,y)=\frac{1}{2}(x^2+y^2)+V_{0}e^{-(x+3)^2/2\sigma^2}+V_{0}e^{-(x-3)^2/2\sigma^2}$ which has two barrier regions where the hidden vortices and corresponding phase defects can occur. The vortex formation in this particular case is very interesting. As shown in Fig. 1(a), in the central potential well, the vortices form a one-dimensional linear chain structure. However, in the two other wells on both side of the central well, the vortex lattice structure is the usual Abrikosov lattice.  Such kind of linear vortex lattice structure is useful in various applications such as quantum computing etc. \cite{Kapale}. The hidden vortices which forms in the two barrier regions around the central potential well are shown as phase defects in the corresponding phase profile in Fig. 1(b). We have also confirmed numerically the  presence of the hidden vortices in this case from the Feynman rule $N_{t}/2=l_{z}$, where $N_t$ is the total number of vortices (visible plus hidden) in the condensate and $l_z$ is the average angular momentum per atom in equilibrium \cite{Feynman,Ueda}.
\par
The Feynman rule for the number of vortices in a superfluid rotating in a rigid container is given by $N=m \Omega R^2/\hbar$, where $R$ is the radius of the container \cite{Feynman}. As mentioned above, the Feynman rule can also be expressed as  $l_z = N/2$. For the harmonically trapped rotating BEC,  the centrifugal force modifies $R$ as $\small R^2(\Omega)=  R_{\bot}(0)^2 (1-\Omega^2)^{-1/2}  $.  From this relation we can see that the radius and hence $N$, diverges when $\Omega\rightarrow 1$. However, no such divergence have been observed in the experiment when $\Omega \rightarrow 1$ \cite{Williams}. Here we derive the Feynman's relation for the rotating BEC when trapped in an anisotropic double-well potential using  the Thomas-Fermi approximation \cite{Fetter}. The effective potential in presence of the rotation is $V_{tot}=\frac{1}{2}m(\omega_x^2 x^2+\omega_y^2 y^2)+V_{0}(1-x^2/2\omega^2)+\frac{1}{2}m\Omega^2(x^2+y^2)$, where we neglected the higher order terms in the expansion of $e^{-x^2/2\omega^2}$. The normalization condition in 2D shows
that $\frac{\mu^{'}(\Omega)}{\mu^{'}(0)}=\bigg(\frac{(1-\frac{\Omega^2}{\omega_x^2}-\frac{V_{0}}{m\omega_x^2 w^2})(1-\frac{\Omega^2}{\omega_y^2})}{(1-\frac{V_{0}}{m\omega_x^2 w^2})}\bigg)^{1/4}$, where $w$ is the width of the barrier. The radius of the condensate can be taken as the average of the condensate radius along x and y directions. That is, $R(\Omega)^2=\frac{1}{2}[R_{x}^2(\Omega)+R_{y}^2(\Omega)]$. Using the similar procedure in reference\cite{Fetter}, we obtain the Feynman rule for the double-well trap potential in our dimensionless form as
$\small N=\Omega R^2(\Omega), $
where $\small R^2(\Omega)=  \frac{\Lambda}{2} \bigg[R_{x}^2(0) (\frac{1-b}{1-b-\Omega^2})+R_{y}^2(0) (\frac{\lambda^2}{\lambda^2-\Omega^2})\bigg],$
 $b=V_{0}/\sigma^2$ and $\Lambda=\big(\frac{(1-\Omega^2-b)(1-\Omega^2/\lambda^2)}{(1-b)}\big)^{1/4}$.
For harmonic symmetric trap b=0, $\lambda=1$, $R_{x}=R_{y}=R_{\bot}$,
$\Lambda=(1-\Omega^2)^{1/2}$
and we get back the result $\small R^2(\Omega)=  R_{\bot}(0)^2 (1-\Omega^2)^{-1/2}$ \cite{Kato}.  
 \begin{figure} [pt]
 \includegraphics[width=8.6cm,height=4cm]{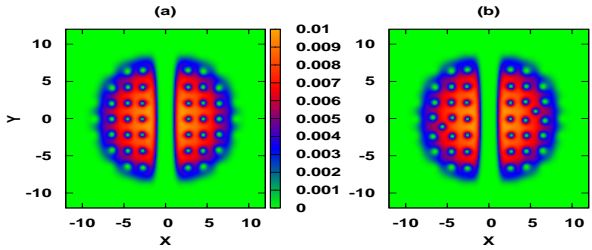}
    \vspace{-20pt}
 \caption{\label{Fig. 4}{\footnotesize (Color online) (a) Pinned vortex lattice for SOL potential at $V_{1}=2.2 $ and (b) at $V_{1}=3 $ and t=250. Here $\Omega=0.9$ and $\sigma=0.5$.}}
    \vspace{-20pt}
 \end{figure}
 From the Feynman formula as derived above, it can be seen that for the isotropic case ($\lambda =1$) the radius of the condensate  $R(\Omega) \rightarrow \infty$  when  $\Omega \rightarrow 1$, similar to the case of harmonic trap.  On the other hand, for the anisotropic case ($\lambda \neq 1$) the radius of the condensate depends on the anisotropy parameter $\lambda$. Whenever $\lambda <1$, the term $\lambda^2 - \Omega^2$ in Eq. (1) becomes negative and the condensate becomes unstable. But for $\lambda >1$, the rotation at $\Omega=1$ is also possible which is in agreement with the experiment.  This is shown in Fig. 2. It can be seen from the figure that the analytical Feynman formula for the number of vortices (Eq. (1)) matches quite well with the total number of vortices $N_t$ as obtained numerically.  The corresponding plot for the isotropic case ($\lambda=1$) is shown in the inset of the figure where we can see that the number of vortices increases at much faster rate when $\Omega \rightarrow 1$ as compared to the anisotropic case.
       \begin{figure}[pb]
         \vspace{-20pt}
 \includegraphics[width=8.6cm,height=4cm]{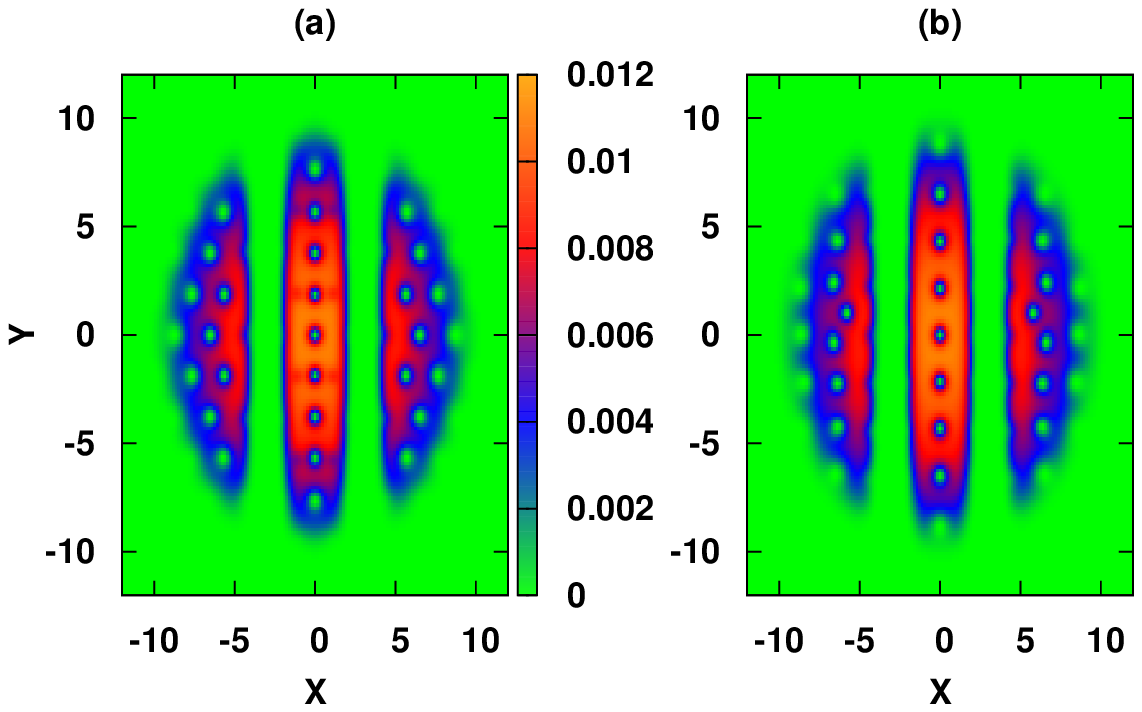}
   \vspace{-25pt}
 \caption{\label{Fig. 5}{\footnotesize (Color online) (a) Pinned vortex lattice for (a) TOL  and (b) SOL potential. Here $p=1000$, $\Omega=0.9$, $V_{1}=3$ and $\sigma=0.5$.}}
 \includegraphics[width=8.6cm,height=4cm]{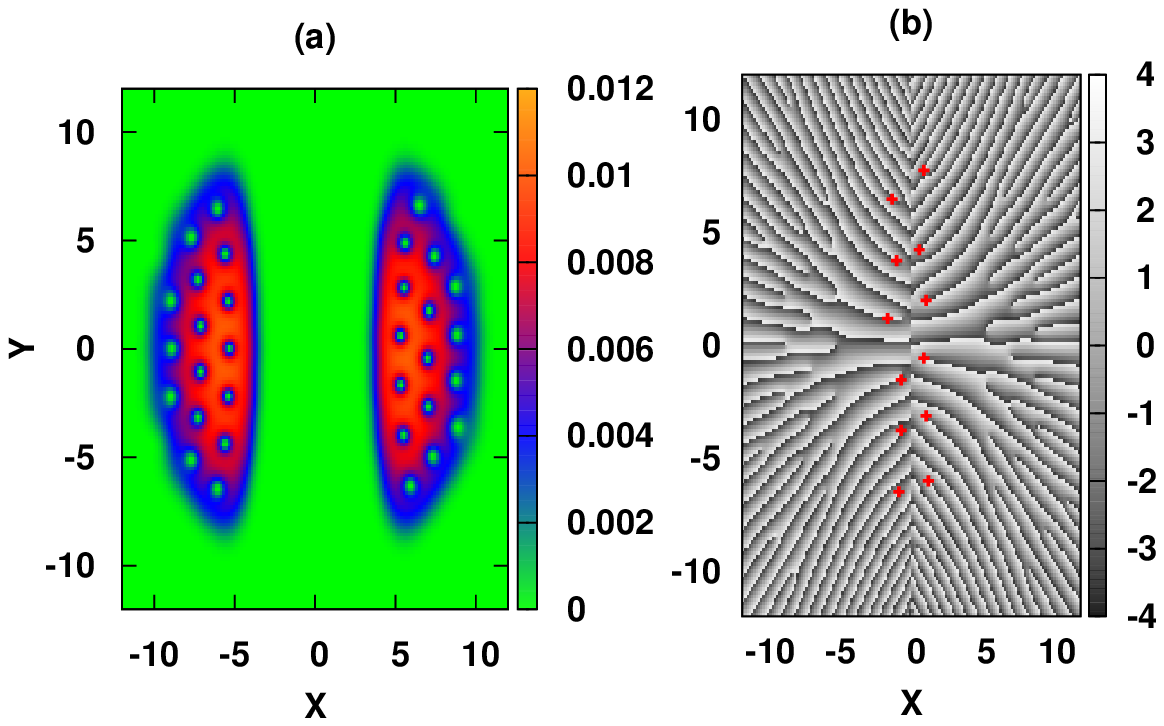}
   \vspace{-25pt}
 \caption{\label{Fig. 6}{\footnotesize (Color online) (a) Condensate density and (b) its phase profile. Here $\Omega=0.9$, $V_{1}=0$ and $\sigma=2$.}}
  \end{figure}
 \par
  In order to study the pinning of vortices, we consider both the square and the triangular co-rotating optical lattice potentials. In Fig. 3 we present the lattice potential energy obtained from numerical simulations for different strength $V_{1}$ of the triangular optical lattice (TOL) (red dots) and square optical lattice (SOL) (blue line-dots). It shows that the transition from Abrikosov vortex lattice to the pinned lattice occur through the intermediate coexisting state, similar to that observed for the case of harmonic trap potential \cite{Sato}.  The comparison between the two cases show that the  strength of the TOL potential required for the pinning of the visible vortices is comparatively lower than that of the SOL case. This is expected since the Abrikosov vortex lattice structure is commensurate with the TOL easily. Fig. 4(a) shows the completely pinned vortex lattice for the SOL with strength $V_1 = 2.2$. When the strength of the SOL is further increased to $V_1 = 3$, we find that there is an extra vortex on both sides of the central barrier region as shown in Fig. 4(b). This two particular defect vortices remains unpinned even if all the other vortices gets pinned.  The defect of this kind has already been observed in experiment for the BECs trapped in a harmonic potential \cite{Tung}. It is interesting to note that such defects do not appear for the TOL (not shown here).  This is because the commensurate nature of the Abrikosov vortex lattice and the TOL which allows all the vortices to get pinned to the optical lattice pinning sites, leaving no room for unpinned defect vortices.

  Fig. 5 shows the pinning of the vortices for the case of rotating BEC in triple-well trap potential. Due to the linear lattice structure of the vortex lattice (see Fig. 1(a)), only alternate  vortex sites in the linear chain of vortices  coincides with the  pinning sites of TOL and can get pinned.  This is exactly seen form our numerical simulation and is shown in Fig. 5(a) where we can see that only alternate vortices of the central linear chain of vortices are pinned. The vortices in the neighbouring two wells form the usual Abrikosov lattice and therefore all these are pinned to the TOL. On the other hand  all the vortex  sites in the linear chain of vortices  coincide with the SOL pinning sites and can get pinned. This is shown in Fig. 5(b) where we can see that all the vortices in the central well are completely pinned. In the neighbouring two wells, all the vortices are also pinned except the defect vortices.
     \begin{figure}[pt]
       \vspace{-25pt}
 \includegraphics[width=8.6cm,height=4cm]{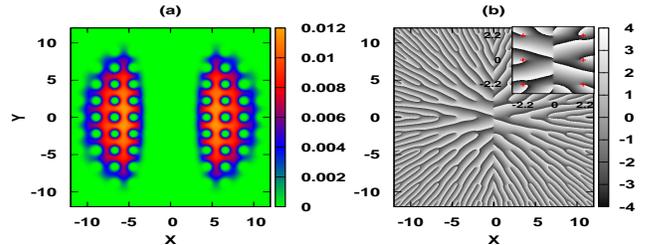}
    \vspace{-23pt}
\caption{\label{Fig. 7}{\footnotesize (Color online)  (a) Pinned vortex lattice for the SOL potential and (b) its phase profile. Here $\Omega=0.9$, $V_{1}=100$ and $\sigma=2$.}}
   \vspace{-15pt}
 \end{figure}
\par
The pinning of the hidden vortices is a difficult problem because the pinning can be seen only as defects in the phase profile of the condensate density. To get a clear picture of the pinning of hidden vortices, we first increase the number of hidden vortices in the condensate. This can be done by increasing the width of the central barrier region of the double-well trap potential large enough (by increasing $\sigma$ ) such that this region can accommodate sufficient number of hidden vortices. Interestingly, by increasing the width, we find the existence of a new type of hidden vortices located in the central barrier region. The phase profile of this new type of hidden vortices is different from the hidden vortices reported earlier in the literature \cite{Wen1}. Fig. 6(a) shows the visible vortices which form the usual Abrikosov lattice on both sides of the larger central barrier region. From the phase profile in Fig. 6(b) we can see that there are two types of such phase defects in the central barrier region. There are some lines which ends in the central barrier line where the phase changes discontinuously and these defects are the the usual hidden vortices reported earlier in the literature. Then, there are other lines where the phase changes discontinuously from black to white (similar to the phase profile of the visible vortices) and the end of these lines represent the phase defects. These phase defects (shown by red dots in Fig. 6(b)) represent a new type of hidden vortices. We identify these phase defects as hidden vortices since there are no visible vortices in these positions (red dots) as seen from Fig. 6(a). This is also checked by calculating the average value of the angular momentum and verifying the  $N_{t}/2=l_z$ rule as mentioned above. From numerical simulations we find that only these new type of hidden vortices gets pinned to the optical lattice sites.  Fig. 7(a) shows the pinning of the visible vortices in SOL potential. Fig. 7(b) shows the corresponding phase profile. As we increase the strength of the SOL potential, the hidden vortices gradually moves away from the central line and finally gets pinned to the SOL sites. The magnified picture of the central region is shown in the right top corner of the figure, where the pinned hidden vortices are shown by red dots. Similar pinning of the hidden vortices is also seen for the TOL for lattice strength $V_1=100$.

 In conclusion, we have derived analytically the surface mode frequency and the Feynman rule for the BEC in a rotating double-well trap potential and compared with numerical results. We have shown the existence of the surface mode with $l=4$ for the double-well potential and the non-divergence of vortex number for the anisotropic double-well potential when $\Omega \rightarrow  1$.  The linear vortex lattice structure as well as its interesting pinning effect for the condensate in a rotating triple-well trap potential is expected to be very useful in applications such as quantum computing. We have shown the existence of a new type of hidden vortices whose phase profile is similar to that of the visible vortices. We have shown that this new type of hidden vortices also gets pinned, but for much higher strength of the optical lattice.

\par
B. D thanks DST for financial support through a research project. K. P. thanks DST, DAE-BRNS, CSIR
and UGC for the financial support through research projects.

\end{document}